\documentclass[%
 reprint,amsmath,amssymb,aps, superscriptaddress
]{revtex4-1}

\usepackage{graphicx,psfrag,float}
\usepackage{dcolumn}
\usepackage{bm}
\usepackage{gensymb}
\newcommand{\eins}{\mbox{$1 \hspace{-1.0mm} {\bf l}$}}
\graphicspath{{Figures3/}}
\begin{document}


\title{Entanglement sudden-death time: a geometric quantity}

\author{Rams\'es J. S\'anchez}
\affiliation{Bethe Center of Theoretical Physics, The University of Bonn, 53115 Bonn, Germany.\\}
\author{Esteban Isasi}
\affiliation{
Departamento de F\'{\i}sica, Secci\'{o}n de Fen\'{o}menos \'{O}pticos, Universidad Sim\'{o}n Bol\'{\i}var, 
apartado postal 89000, Caracas 1080A, Venezuela.\\
}
\author{Douglas Mundarain}
\affiliation{
Departamento de F\'{\i}sica, Secci\'{o}n de Fen\'{o}menos \'{O}pticos, Universidad Sim\'{o}n Bol\'{\i}var, 
apartado postal 89000, Caracas 1080A, Venezuela.\\
}

\date{December 15, 2014}

\begin{abstract}
We study the entanglement evolution of the set of Bell diagonal states for a two-qubit system coupled to two independent vacuum noise sources. This set can be represented geometrically as the set of points inside a tetrahedron in a three-dimensional Euclidean space and contains the maximally entangled states for bipartite systems. We show that the set of entangled Bell diagonal states can be divided into two bounded subsets in this representation: states that evolve into separable states in a finite time and states that lose their entanglement asymptotically. Additionally, we find that the finite time in which the Bell diagonal states lose their entanglement depends only on the distances from their position in the three-dimensional representation to the boundaries of both, the set of separable states and the set of states that remains always entangled.
\end{abstract}

\maketitle


\section{Introduction}

Entanglement can be understood as the emergence of non-local correlations among physical systems that have interacted, and it originates as a consequence of the superposition principle inherent to the Hilbert space of the composite system \cite{Schrodinger1,Schrodinger2}. These non-local correlations can be thought of as a resource that can be used to design protocols for quantum communication and quantum computation  \cite{Nature.402.390, Nature.404.247, PhysRevLett.86.5188,PhysRevLett.68.3121,PhysRevLett.69.2881,PhysRevLett.70.1895}. However, any implementation of such protocols inevitably involves the coupling of the system of interest to the uncontrolled degrees of freedom of an environment. This coupling always induces the decay of classical and quantum correlations alike and, consequently, the loss of entanglement \cite{ZPhysB.59.223,PhysRevA.69.052105}. Remarkably, while the local capability of each part of the composite system to show interference is always suppressed asymptotically \cite{Diosi}
, the 
entanglement between these parts can either vanish completely in a finite time, or decay exponentially. These two different kinds of disentanglement dynamics are known as entanglement sudden death (ESD) and entanglement asymptotic death (EAD), respectively. Both behaviors were originally predicted for a pair of non-interacting qubits coupled to two uncorrelated vacuum noise sources \cite{PhysRevLett.93.140404, PhysRevLett.97.140403}. Since then, ESD and EAD have been investigated for two-qubit systems immersed in correlated reservoirs \cite{PhysRevA.74.024304}, multiple-qubit systems \cite{PhysRevLett.101.080503} and spin chains \cite{PhysRevA.78.012357}, to name a few models, and have been observed experimentally \cite{Science.316.579, PhysRevLett.99.180504}. 

Generally speaking, whether the entanglement of an open system will fade away completely in a finite or infinite time depends on both, the kind of coupling to the environment and the initial degree of entanglement among the parts of that system \cite{Science.323.598}. Since any quantum information technology largely relies on the preservation of the entanglement of a system open to background noise, it is of interest to control which kind of long-time behavior a particular initial preparation of this system will end up with. Furthermore, it is known that local operations and classical communication (LOCC) between bipartite systems can be used to partially recuperate the entanglement lost to the environment, provided the system is entangled at the time the LOCC are implemented \cite{PhysRevLett.78.574}. This means that LOCC will not succeed when applied to a system showing ESD, after its entanglement has vanished completely. Therefore, given an initial preparation of a system, it is of practical relevance to 
know the time that will take its entanglement to disappear. We will refer to this time as entanglement sudden-death time (SDT) henceforth. Thus, a classification of the initial states of an entangled system in terms of their SDT can be useful and, as we will see, a description of the state space in geometrical terms can be helpful in obtaining such a classification. In fact, it is possible to understand the phenomena of ESD and EAD in terms of the position and geometry of the asymptotic states into which the system evolves, relative to the set of separable states \cite{TerraOCunha2007,TerraOCunha2009}. This general result, however, is not directly related to the geometry of the set of initial conditions. 

For two-qubit systems, an alternative geometrical description of the state space is given in terms of equivalence classes. We shall consider the two states described by the density matrices $\rho$ and $\tilde{\rho}$ as equivalent, provided 

\begin{equation}
 \tilde{\rho} = M \rho M^{\dagger},
\label{eq:equivalence}
\end{equation}

\noindent
 with $M= M_1 \otimes M_2$ and $M_{1,2} \in SL(2,\mathbb{C})$, the group of $2 \times 2$ matrices with complex entries and unit determinant \cite{VerFDD2001,PhysRevA.74.012313, AvronKenethANNPhys2009}. These operations transform entangled (separable) states into entangled (separable) states with finite success probability and hence are a special case of LOCC. This implies that when studying the disentanglement dynamics of a bipartite open system, all initial two-qubit states can be filtered down to their $SL(2,\mathbb{C})$ representatives and be labeled by the SDT of this reduced set of states. Consequently, a classification of the class representatives in terms of the SDT gives an indirect classification to the full set of initial conditions for an open two-qubit system. What is more, such a classification has an additional appeal, namely that the set of $SL(2,\mathbb{C})$ representatives contains the maximally entangled states that can be obtained through any LOCC  \cite{PhysRevLett.83.2656}.  In this work 
we present a SDT classification of the $SL(2,\mathbb{C})$ representatives, as the set of initial conditions for a pair of non-interacting qubits, coupled locally to two independent cavities at zero temperature. We find that the set of representatives with finite entanglement can be divided into two bounded subsets of states, each of which evolves into states exhibiting either ESD or EAD. Additionally, we show that the SDT of the  normalized representatives which undergo ESD depends only on geometric quantities.

\section{Geometric picture of the $SL(2, \mathbb{C})$ representatives}

In the basis of the Pauli matrices $\{ \sigma_i \otimes \sigma_j \}$ $(i,j=0,1,2,3)$ chosen as

\begin{equation}
\begin{array}{c}
\displaystyle \sigma_0= \left( \begin{array}{cc}
	        1 & \phantom{-}0 \\
	        0 & \phantom{-}1
	        \end{array} \right)   \qquad \sigma_1 = \left( \begin{array}{cc}
	                                                0 & \phantom{-}1 \\
	                                                1 & \phantom{-}0
	                                               \end{array} \right)\phantom{,} \\ \\  \sigma_2 = \left( \begin{array}{cc}
	                                                                                       0 & -i \\
	                                                                                       i &  \phantom{-}0
	                                                                                      \end{array} \right) \qquad \sigma_3 = \left( \begin{array}{cc}
	                                                1 &  \phantom{-}0 \\
	                                                0 & -1
	                                               \end{array} \right),
\end{array}
\end{equation}

\noindent
any two-qubit state $\rho$ can be represented by the real, $4\times4$ matrix $R$ as follows

\begin{equation}
 \rho = \displaystyle \frac{1}{4} \sum_{i,j = 0}^3 R_{ij} \sigma_i \otimes \sigma_j.
 \label{eq:rho}
\end{equation}

The $SL(2,\mathbb{C})$ operations in (\ref{eq:equivalence}) define four equivalence classes and give a certain classification to the state space of two qubits. These operations act on $R$ by a pair of proper, orthochronous Lorentz transformations $L_1$ and $L_2$ such that $\tilde{R}=L_1^{\phantom T}R L_2^{T}$ is the representative of the equivalence class \cite{VerFDD2001}. The first class consists of the set of states which transforms into the diagonal representative
\begin{equation}
 \tilde{R}=\mbox{diag}(X_0,X_1,X_2,X_3),
 \label{eq:regular}
\end{equation}

\noindent
where $X_0 \ge |X_i|$ $(i=1,2,3)$.  In the literature, the real numbers $X_i$ $(i=0,1,2,3)$ are known as the Lorentz singular values of the matrix $R$. These numbers allow to depict the diagonal representatives as  the set of points forming the interior of a four-dimensional convex cone. The second equivalence class corresponds to the states which transform into the non-diagonal representative

\begin{equation}
 \tilde{R}=\left( \begin{array}{cccc}
                   X_0+k & 0 & 0 & -k \\
                   0 &  X_1 & 0 & 0 \\
                   0 & 0 & X_2 & 0 \\
                   k & 0 & 0 & X_3-k
                  \end{array} \right),
\label{eq:irregular}
\end{equation}

\noindent
with $X_0=X_3 \ge |X_{1,2}|$, $X_1+X_2=0$ and $k$ fixed and positive. This set can then be interpreted as forming the boundary of the four dimensional cone. Note that, since normalization implies $\mbox{Tr}(\tilde{\rho})=\tilde{R}_{00}=1$, not all these representatives are normalized. This is due to the fact that (\ref{eq:equivalence}) does not preserve the normalization of the original density matrix. Finally, the last two equivalence classes have vanishing Lorentz singular values and, accordingly, are represented by the apex of the cone. Unlike the set of representatives in (\ref{eq:regular}) and (\ref{eq:irregular}), these last two equivalence classes consist of separable states only and their SDT classification is trivial. 


\begin{figure}
\includegraphics[width=6.0cm]{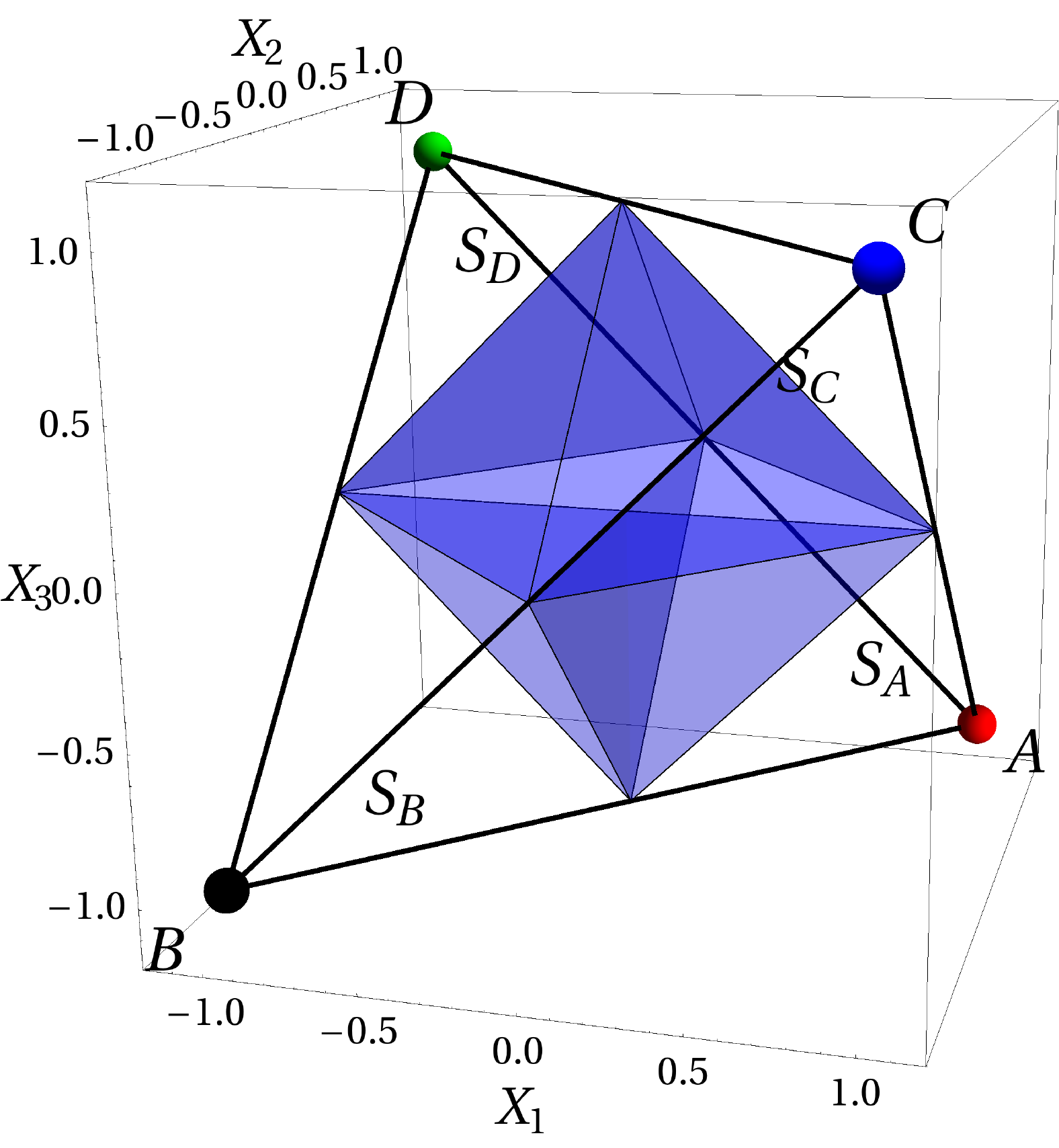}
\caption{Geometric representation of the set of Bell diagonal states. The octahedron contains the set of separable states. The points outside the octahedron and inside the tetrahedron represent the entangled states. The points $A$, $B$, $C$ and $D$ label the four Bell states. In particular, $B$ is the representative of the singlet state $|B \rangle=\frac{1}{\sqrt{2}}\left(|01\rangle-|10\rangle\right)$.}  
\label{fig:tetrahedron}
\end{figure}

 The subset of normalized states in equation (\ref{eq:regular}) is particularly interesting because its entangled states are the maximally entangled states that can be obtained through any LOCC  \cite{PhysRevLett.83.2656}. This set is diagonal in the basis of Bell states \cite{PhysRevLett.68.3259} and corresponds to a cross section of the four dimensional cone at $X_0=1$, thereby forming a three-dimensional tetrahedron \cite{PhysRevA.74.012313}. The tetrahedron is inscribed in a cube of size two, centered at the origin of the coordinate system. The $X_i$ $(i=1,2,3)$ are then identified as the coordinates of the points inside the tetrahedron. In particular, its vertices are given by the points

\begin{equation}
\arraycolsep=1.4pt\def\arraystretch{1.5}
 \begin{array}{cc}
   A=(1,\phantom{-}1,-1) & \quad   B=(-1,-1,-1)\phantom{,} \\ 
   C=(1,-1,\phantom{-}1) &  \quad D=(-1,\phantom{-}1,\phantom{-}1),
 \end{array}
\label{eq:vertices}
\end{equation}

\noindent
and correspond to the four Bell states. The set of Bell diagonal states contains a subset of separable states which, in turn, forms an octahedron whose corners correspond to the middle points on the edges of the tetrahedron \cite{PhysRevA.74.012313}. The faces of the octahedron are known as Peres-Horodecki planes. It follows that the entangled Bell diagonal states are the set of points outside the octahedron and inside the tetrahedron. These states then form four disconnected pyramids which we label as $S_A$, $ S_B$, $S_C$ and $S_D$, according to the vertex of the tetrahedron they contain (see Fig. \ref{fig:tetrahedron}). 

\section{Model}

Let us consider a system of two entangled qubits, coupled individually to two independent cavities at zero temperature. Each cavity is treated as a large set of coupled harmonic oscillators with vanishing average photon number and vanishing correlation time. Hence, each qubit is under the sole effect of vacuum fluctuations. The coupling between the harmonic oscillators and the qubits is described within the dipolar and the rotating-wave approximation \cite{OrszagBook}. In order to study the disentanglement dynamics of this system, we calculate the time evolution of its density matrix in the Born-Markov limit and quantify its entanglement through the concurrence \cite{PhysRevLett.80.2245}. The time evolution of the density matrix $\rho$ is then governed by the master equation 

{\small 
\begin{equation}
\begin{array}{rcl}
\displaystyle \frac{d \rho(t)}{dt} & = & \displaystyle \frac{\gamma}{2}\left[ \left(2\sigma^{(1)} \rho {\sigma^{(1)}}^{\dagger}-{\sigma^{(1)}}^{\dagger} \sigma^{(1)} \rho -\rho {\sigma^{(1)}}^{\dagger} \sigma^{(1)}\right) \right. \\ \\
 & & + \displaystyle \left. \left(2\sigma^{(2)} \rho {\sigma^{(2)}}^{\dagger}-{\sigma^{(2)}}^{\dagger} \sigma^{(2)} \rho -\rho {\sigma^{(2)}}^{\dagger} \sigma^{(2)}\right) \right],
\end{array}
\label{eq:master}
\end{equation}
}

\noindent
where $\gamma$ is the vacuum decay constant and 

\begin{equation}
 \sigma^{(1)}=\sigma \otimes \eins, \quad \sigma^{(2)}= \eins \otimes \sigma, \quad \sigma^{\phantom\dagger}=\left( \begin{array}{cc} 
                0 & 0 \\
                1 & 0 
               \end{array} \right).  
\end{equation}

We are interested in studying the time evolution of the $SL(2,\mathbb{C})$ representatives (\ref{eq:regular}) and (\ref{eq:irregular}) under the influence of the two uncorrelated baths. These states are particular cases of a family of states known as {\it X-states} \cite{JPA.41.412002} whose $R$ matrix reads

 \begin{equation}
 R=\left( \begin{array}{cccc}
                   x_0 & 0 & 0 & x_4 \\
                   0 &  x_1 & 0 & 0 \\
                   0 & 0 & x_2 & 0 \\
                   x_5 & 0 & 0 & x_3
                  \end{array} \right)
\label{eq:Xstate}
\end{equation}

We consider this set of states because it maintains its matrix structure when evolving with the master equation (\ref{eq:master}). This allows us to obtain an analytic expression for the time-depend concurrence. Solving equation (\ref{eq:master}), for the set of states in (\ref{eq:Xstate}), one obtains

\begin{equation}
 x_0(t)=X_0,
 \label{eq:0}
\end{equation}
\begin{equation}
 x_{(1,2)}(t)=X_{(1,2)} e^{-t \gamma},
 \label{eq:1}
\end{equation}

\begin{equation}
\arraycolsep=1.4pt\def\arraystretch{1.8}
\begin{array}{rcl}
 \displaystyle x_3(t) & = & e^{-2t \gamma} \Big\{ X_0 \left( e^{t \gamma}-1\right)^2+X_3  \\
 \displaystyle        &   &  -\left(X_4+X_5\right)\left( e^{t \gamma}-1\right) \Big\},
\end{array}
 \label{eq:3}
\end{equation}
\begin{equation}
 x_{(4,5)}(t)=X_0\left(e^{-t \gamma}-1\right)+X_{(4,5)}e^{-t \gamma},
 \label{eq:4}
\end{equation}

\noindent
with the initial conditions $x_i(t=0)=X_i$, $(i=0, 1, ... ,5)$. The entanglement evolution of the two-qubit states is then studied through the concurrence, defined as \cite{PhysRevLett.80.2245}

\begin{equation}
 C(\rho)=\mbox{max}\left\{ 0, \, 2 \sqrt{\lambda_{\mbox{\tiny max}}}-\sum_{i=0}^3\sqrt{\lambda_i} \right\},
 \label{eq:concurrence}
\end{equation}

\noindent
where the $\lambda_i$ $(i=0,1,2,3)$ are the eigenvalues of the matrix $\hat{\rho}=\rho(\sigma_2\otimes \sigma_2) \rho^{*} (\sigma_2\otimes \sigma_2)$, with $\rho^*$ the complex conjugated of the density matrix $\rho$.

\subsection{Classification of the Bell diagonal states}

We start the classification of the class representatives by considering the set of Bell diagonal states as the set of initial conditions for our problem. To choose these states as initial conditions amounts to set $X_0=1$ and $X_4=X_5=0$ in equations (\ref{eq:0}-\ref{eq:4}). The $X_i$ $(i=1,2,3)$ are identified with the coordinates of the set of points in the tetrahedron of Fig. \ref{fig:tetrahedron}. Using definition (\ref{eq:concurrence}) one can show that, for this set of initial conditions, the concurrence is given by

\begin{equation}
\small
\begin{array}{rcl}
 C_1(t,\vec{X}) & = & \displaystyle \frac{1}{2}\Bigr( e^{-t \gamma} |X_1+X_2|- e^{-2t \gamma} \sqrt{\left(1+X_3\right)} \\ \\
                &   & \displaystyle \times \sqrt{1-4e^{t \gamma}+4e^{2t \gamma}+X_3 } \Bigr),
\end{array}
\label{eq:C1}
\end{equation}

\noindent
if $X_3 < 0$. Otherwise the concurrence is 

\begin{equation}
\small
C_2(t,\vec{X})=\frac{1}{2}\left( e^{-t \gamma} |X_1-X_2|+e^{-2t \gamma}\left(1-2e^{t \gamma}+X_3\right) \right)
\label{eq:C2}
\end{equation}

 \begin{figure*}[]
 \centering
 \begin{minipage}[b]{7.0cm}
\centering
\includegraphics[width=6.0cm]{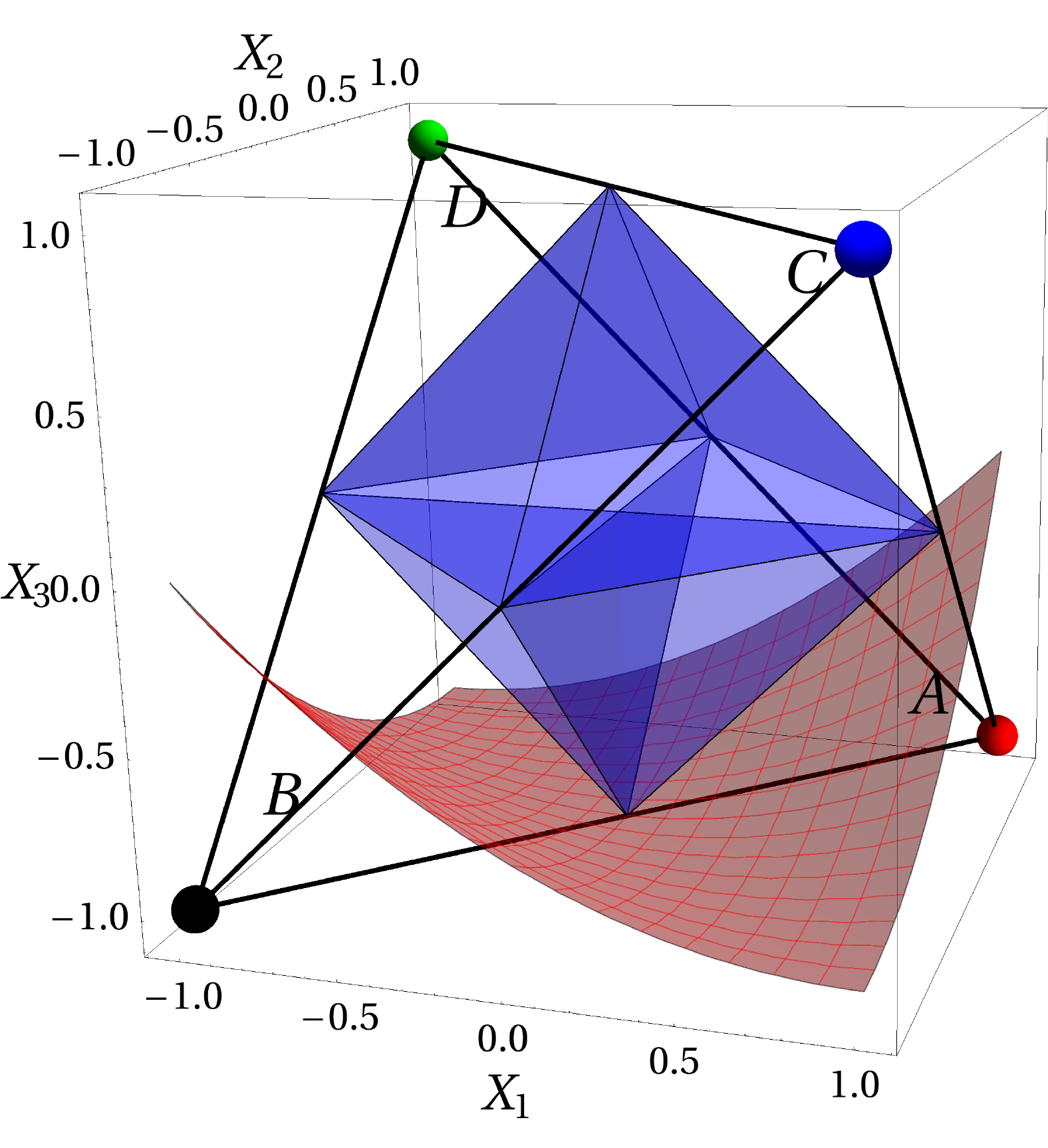}
\end{minipage}
\begin{minipage}[b]{7.0cm}
\centering
\includegraphics[width=6.5cm]{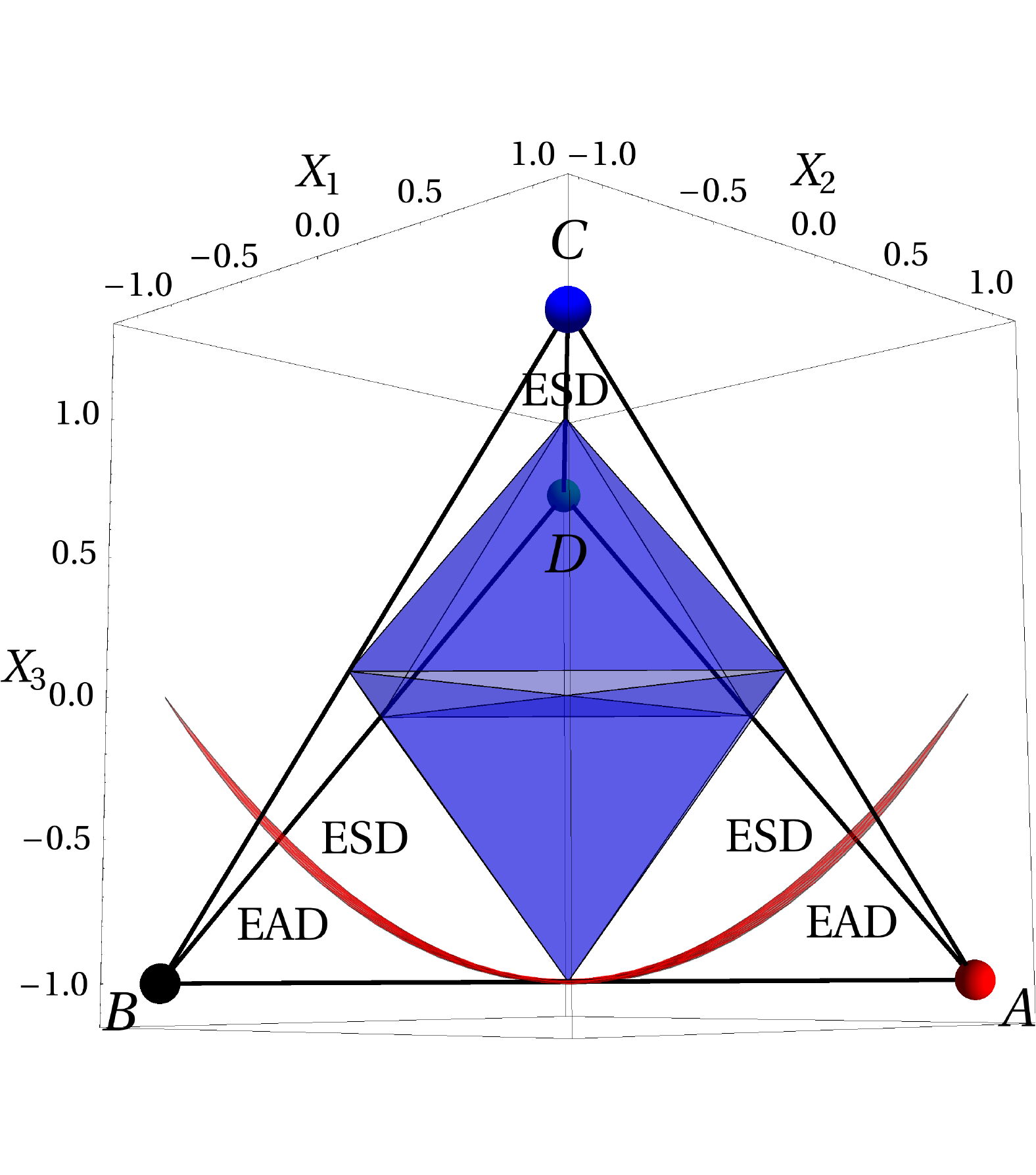}
\end{minipage}
\caption{Geometric classification of the set of Bell diagonal states. The points inside the octahedron correspond to separable states. The quadratic surface divides the initial conditions which evolve into separable state in a finite time (ESD) from those which evolve into separable states asymptotically (EAD)}
 \label{fig:regions}
\end{figure*} 

Geometrically, $C_1$ describes the concurrence for the set of states whose initial conditions lie in the regions of the tetrahedron given by the set of points $S_A \cup S_B$. Likewise, $C_2$ describes the concurrence for the states whose initial conditions are represented by the set $S_C \cup S_D$. From these expressions we observe that, as was pointed out in \cite{PhysRevLett.93.140404}, one can find initial states that show either ESD or EAD. An example of the former case is given by the state $\alpha=(-0.5,-0.7,-0.3)$ for which $C_1(t)$ vanishes exactly when $t/\gamma \approx 0.624$. The singlet state $B=(-1,-1,-1)$, in contrast, has $C_1(t)=e^{-t \gamma}$ and then shows EAD.

In order to classify the entire set of entangled Bell diagonal states $(S_A \cup S_B \cup S_C \cup S_D)$ in subsets whose evolution leads to states exhibiting either ESD or EAD, we have to identify the initial conditions for which the concurrence vanishes in a finite time. Let us look first at the states that lie in $S_A \cup S_B$, for which the concurrence is given by $C_1(t)$. If we set $C_1(t)$ to zero and make the change of variables $y=e^{t\gamma} > 1$, we obtain 

\begin{equation}
(1+X_3)(1+X_3-4y+4y^2)-y^2(X_1+X_2)^2= 0.
 \label{eq:condition1}
\end{equation}

This equation is fulfilled by states in $S_A \cup S_B$ provided

\begin{equation}
 \frac{1}{4}\left(X_1+X_2\right)^2-1 < X_3 < -1+|X_1|+|X_2|,
 \label{eq:inequality_2}
\end{equation}

\noindent 
where the lower bound corresponds to the long-time limit $(y \gg 1)$ of equation (\ref{eq:condition1}), whereas the upper bound corresponds to the limit of vanishing time $(y \rightarrow 1)$. The resulting inequalities, together with the boundaries of the tetrahedron, define two regions $S_A' \subseteq S_A$ and $S_B' \subseteq S_B$ which, by construction, contain a set of Bell diagonal states for which the concurrence vanishes in a finite time. The boundaries of these regions are given by the faces of the tetrahedron, the Peres-Horodecki plane defined here by

\begin{equation}
 P(\vec{X})=-1+|X_1|+|X_2|-X_3=0,
 \label{eq:plane}
\end{equation}

\noindent
and the quadratic surface

\begin{equation}
Q(\vec{X})=\frac{1}{4}\left( X_1+X_2\right)^2-X_3-1=0.
\label{eq:surface}
\end{equation}

Note that both surfaces are open boundaries, since their points do not belong to $S_A' \cup S_B'$. Indeed, the states on the quadratic surfaces disentangle asymptotically while the states on the plane are separable for all times. Therefore, every initial state in $S_A' \cup S_B'$ evolves into a state with ESD, whereas the initial entangled states that do not belong to this region, evolve into states with EAD. Both sets are shown in Figure \ref{fig:regions}. 

A similar analysis shows that for the region $S_C \cup S_D$, the Bell diagonal states whose concurrence vanishes in a finite time are such that

\begin{equation}
 y|X_1-X_2|+(1-2y+X_3)=0,
 \label{eq:condition2}
\end{equation}

\noindent
with $y = e^{t \gamma} \ge 1$. Here, the long-time limit of equation (\ref{eq:condition2}) yields the two planes $|X_1-X_2|=2$ which intercept the tetrahedron only at the points $C$ and $D$. Hence, in $S_C \cup S_D$ only the Bell states show EAD. The rest of the states in this set becomes separable in a finite time. 

These results extend naturally to the four-dimensional picture of the (non-normalized) diagonal representative, simply by replacing $X_i \rightarrow X_i/X_0$, $(i=1,2,3)$. Thus, we obtain that the convex cone defined by the four Lorentz singular values is cut into two subsets bounded by the $4$-surfaces

\begin{equation}
 X_0=|X_1|+|X_2|+|X_3|,
 \label{eq:4separable}
\end{equation}
\begin{equation}
 X_0+X_3=\frac{(X_1+X_2)^2}{4 X_0}.
 \label{eq:4surface}
\end{equation}

These surfaces define, respectively, the convex cone of separable states and a convex region inside the convex cone of diagonal representatives, whose outer boundary is given by equation (\ref{eq:4surface}).

\subsection{Classification of the non-diagonal representative}

We turn now our attention to the non-diagonal representative (\ref{eq:irregular}) as the set of initial conditions for our model. Setting $X_0=X_0+k$, $X_1=-X_2$, $X_3=X_0-k$ and $X_5=-X_4=k>0$ in (\ref{eq:0}-\ref{eq:4}), and using definition (\ref{eq:concurrence}) for the concurrence, yields

\begin{equation}
\begin{array}{rcl}
 C(t,X_0,X_1) & = & e^{-t \gamma} |X_1|-e^{-2t \gamma} \sqrt{X_0(e^{t \gamma}-1)} \\ \\
              &   & \times \sqrt{(-X_0+e^{t \gamma}(2k+X_0))}.
\end{array}
\end{equation}

As before, we make the change of variables $y=e^{t \gamma} \ge 1$ and set $C=0$ to get

\begin{equation}
 y^2 X_1^2-X_0(y-1)\left\{y(2k+X_0)-X_0\right\}=0.
 \label{eq:condition3}
\end{equation}

By definition $X_0\ge0$ and hence equation (\ref{eq:condition3}) is fulfilled by non-diagonal representatives if

\begin{equation}
 X_1^2 < 2k X_0+X_0^2.
 \label{eq:inequality_3}
\end{equation}

However, the non-diagonal representatives are such that $X_0 \ge |X_1|$, which means that all these states satisfy (\ref{eq:inequality_3}). Therefore, all non-diagonal representatives evolve into states with ESD.

\section{SDT: a geometric quantity}

\begin{figure}
\includegraphics[width=5.0cm]{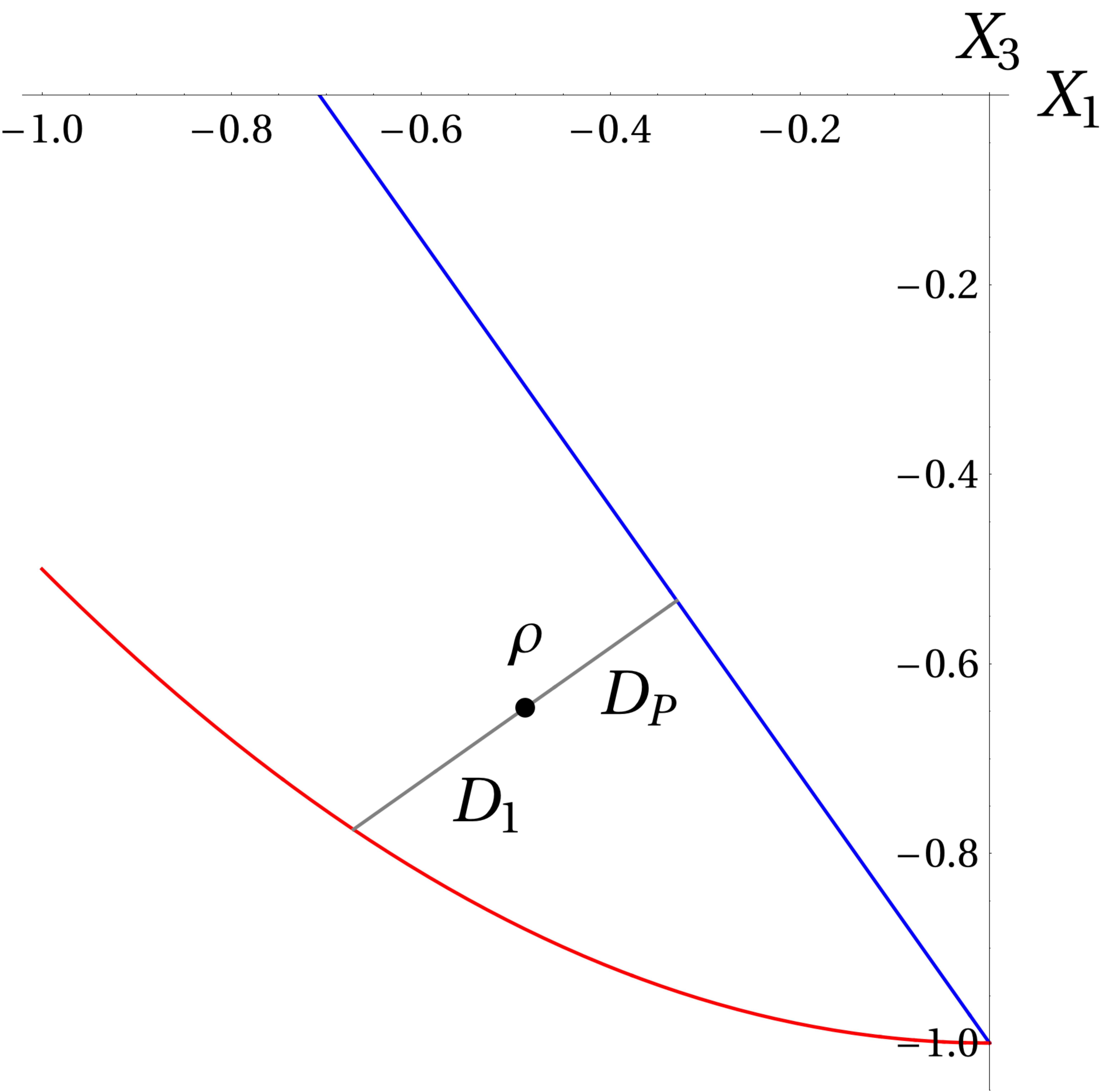}
\caption{Quadrant containing the singlet state $B$ (here, $X_1$ has been rotated $45\degree$ counterclockwise about $X_3$). The straight line intercepting both axes represents the Peres-Horodecki plane (\ref{eq:plane}) and the curve represents the quadratic surface (\ref{eq:surface}). The point $\rho$ corresponds to an entangled Bell diagonal states located at a distance $D_P$ from the plane. The positive quantity $D_1$ is defined in (\ref{eq:DS}).}  
\label{fig:distancias}
\end{figure}

In this section we focus only on the set of Bell diagonal states which exhibit ESD. We start by considering those states lying in the region $S'_A\cup S'_B$ which is defined by equation (\ref{eq:inequality_2}). The finite time that take these states to disentangle can be written down explicitly as a function of their coordinates from the solution of equation (\ref{eq:condition1}) as

\begin{equation}
 t=\frac{1}{\gamma} \mbox{Log} \left[ \frac{-(1+X_3)}{\sqrt{(X_1+X_2)^2-4 X_3}-2}\right].
 \label{eq:time}
\end{equation}

Let us consider for the moment a state in $S'_B$. The Peres-Horodecki plane (\ref{eq:plane}) in this quadrant represents the frontier from which the states are separable for all times. On the other hand, the quadratic surface (\ref{eq:surface}) represents the frontier from which the states lose their entanglement only asymptotically. This suggests that the states close to the plane and far from the quadratic surface evolve into separable states faster than those states close to the surface. Hence, one can intuitively expect that the STD of a state in $S'_B$ depends on the distances from its position to both, the plane and the quadratic surface. In order to show that this is indeed the case, we consider the distance from an initial state $(X_1,X_2,X_3)$ in $S'_A \cup S'_B$, to the Peres-Horodecki plane of the respective quadrant. This distance can be written as

\begin{equation}
 D_{P}=-1+|X_1|+|X_2|+|X_3|.
\end{equation}

We include the dependence on the distance to the surface through the quantity 

\begin{equation}
 D_{1}=1+X_3-\frac{1}{4}(X_1+X_2)^2,
 \label{eq:DS}
\end{equation}

\noindent
 which corresponds to the length of the line perpendicular to the Peres-Horodecki plane that goes from the point $(X_1,X_2,X_3)$ to the quadratic surface (see Fig. \ref{fig:distancias}). Using these two quantities, one can rewrite equation (\ref{eq:time}) as

\begin{equation}
 t=\frac{1}{\gamma} \mbox{Log} \left[\frac{\frac{1}{2}D_P-1+\sqrt{1-D_P-D_1}}{\sqrt{1-D_1}-1}\right].
 \label{eq:SDT1}
\end{equation}

From this result it is clear that the SDT increases as one approaches the quadratic surface. The limit values can be calculated directly 
\begin{equation}
 \displaystyle \lim_{D_1 \to 0} t \, [D_1,D_P]=\infty,
\end{equation}
\begin{equation}
 \lim_{D_P \to 0} t \, [D_1,D_P]= 0.
\end{equation}

Similarly, one can show that the SDT of the initial states in $S_C \cup S_D$ can be written as

\begin{equation}
 t=\frac{1}{\gamma} \mbox{Log} \left[\frac{D_P}{D_2}+1\right],
 \label{eq:SDT2}
\end{equation}

\noindent
with $D_2 = 2-|X_1|-|X_2|$ the length of the line perpendicular to the Peres-Horodecki plane that goes from the point $(X_1,X_2,X_3)$ to the plane $|X_1-X_2|=2$. These two planes play the same role as the quadratic surface in $S_A \cup S_B$, namely that the points on $|X_1-X_2|=2$ disentangle only asymptotically. However, these planes intercept the tetrahedron only at the points $C$ and $D$.

We remark that it is also possible to write the SDT in either of equations (\ref{eq:SDT1}) or (\ref{eq:SDT2}) as a function of the usual Euclidean distances to the quadratic surface, or to the planes $|X_1-X_2|=2$, instead of as a function of $D_1$ and $D_2$. Nonetheless, the resulting expressions are more complicated and their physical content is the same. 

The interesting point about these results is that, in the four-dimensional representation, the concurrence of each representative is nothing but the distance from the point $(X_0,X_1,X_2,X_3)$ that labels it, to the boundary of the set of separable states \cite{AvronKenethANNPhys2009}. This feature is recovered in the three-dimensional representation, as can be readily seen by setting $t=0$ in equations (\ref{eq:C1}) and (\ref{eq:C2}). Similarly, the SDT of the $SL(2,\mathbb{C})$ representatives evolving under the influence of two uncorrelated baths depends on both, the distance from their position to the boundary of the set of separable states, i.e. the initial concurrence, {\it and} the distance from their position to the boundary of the set of representatives that exhibit EAD.

\section{Conclusions}

We studied the entanglement dynamics of a two-qubit system, coupled to two independent thermal baths at zero temperature. We showed that it is possible to give an indirect classification to the full set of initial conditions, in terms of the SDT of the representatives of the four $SL(2,\mathbb{C})$ equivalence classes, defined by transformation (\ref{eq:equivalence}). Here, it is understood that the states that evolve with EAD have infinite SDT. We found that the set of entangled diagonal representatives can be divided into two bounded subsets, each of which evolves into states with either ESD or EAD. In particular, the normalized diagonal representative with $X_3 \le \frac{1}{4}(X_1+X_2)^2-1$ and the two Bell states labeled by the points $C$ and $D$ in equation (\ref{eq:vertices}), remain entangled for all times. As regards the non-diagonal representatives, we found that all of them evolve into states with ESD.

Note nonetheless that this classification does not propagate through the class. In fact, if one considers the state 
\begin{equation}
 R=\left( \begin{array}{cccc}
                   1 & 0 & 0 & -0.24 \\
                   0 & -0.25  & 0 & 0 \\
                   0 & 0 & -0.25 & 0 \\
                   -0.24 & 0 & 0 & -0.5
                  \end{array} \right),
\end{equation}

\noindent
one can calculate its class representative, following \cite{IsasiMundarainStephany2008}, to be 
\begin{equation}
 \tilde{R}=\mbox{diag}(1,-0.304878,-0.304878,-0.829268).
\end{equation}

According the the classification presented here, the class representative $\tilde{R}$ belongs to the set of states that evolves with ESD. However, the original state does not separate in a finite time. 

Finally, we observed that the appearance of ESD and EAD is related to the distance from the representative to the boundary of the set of separable states, in a similar way as the entanglement of the representative is related to this same distance. Specifically, we showed that the finite time which takes the entanglement of the representatives to dissapear depends only on their position with respect to both, the boundary of the set of separable states and the boundary of the set of states that evolves with EAD.

\begin{acknowledgments}
This research used resources from the Centro de C\'omputo
Cient\'{\i}fico del grupo de Relatividad y Campos from the Sim\'on
Bol\'{\i}var University.
\end{acknowledgments}

\bibliography{bibliography}

\end{document}